\documentclass{emulateapj}
\usepackage{epstopdf}

\slugcomment{accepted to AJ}

\shorttitle{Oosterhoff effect in NGC$\,$1851}
\shortauthors{Kunder et al.}

\newcommand{\be}{\begin{equation}}
\newcommand{\ee}{\end{equation}}

\def\c2{\chi ^2}

\begin{document}

\title{The Horizontal Branch of NGC$\,$1851: Constraints from its RR Lyrae Variables}

\author{Andrea Kunder\altaffilmark{1},
Maurizio Salaris\altaffilmark{2},
Santi Cassisi\altaffilmark{3}, 
Roberto de Propris\altaffilmark{1},
Alistair Walker\altaffilmark{1},
Peter B. Stetson\altaffilmark{4},
M\'{a}rcio Catelan\altaffilmark{5,6},
\and
P\'{i}a Amigo\altaffilmark{5,6}
}

\altaffiltext{1}{NOAO-Cerro Tololo Inter-American Observatory, Casilla 603, La Serena, Chile}
\affil{E-mail: akunder@ctio.noao.edu}
\altaffiltext{2}{Astrophysics Research Institute, Liverpool John Moores University, Twelve Quays House, Egerton Wharf, Birkenhead CH41 1LD, UK}
\altaffiltext{3}{INAF-Osservatorio Astronomico di Collurania, Via M. Maggini, I-64100 Teramo, Italy}
\altaffiltext{4}{ Dominion Astrophysical Observatory, Herzberg Institute of Astrophysics, National Research Council, Victoria BC, Canada}
\altaffiltext{5}{Pontificia Universidad Cat\'olica de Chile, Departamento de Astronom\'\i a y Astrof\'\i sica, Av. Vicu\~{n}a Mackenna 4860, 782-0436 Macul, Santiago, Chile; e-mail: mcatelan@astro.puc.cl}  
\altaffiltext{6}{The Milky Way Millennium Nucleus, Av. Vicu\~{n}a Mackenna 4860, 782-0436 Macul, Santiago, Chile}

\begin{abstract}
We use the pulsational properties of the RR Lyrae variables in the globular cluster NGC$\,$1851 
to obtain detailed constraints of the various sub-stellar populations present along
its horizontal branch.  On the basis of detailed synthetic 
horizontal branch modeling, we find that minor helium variations ($Y$$\sim$0.248-0.280) 
are able to reproduce the observed periods and 
amplitudes of the RR Lyrae variables, as well as the frequency of fundamental and 
first-overtone RR Lyrae stars.  
Comparison of number ratios amongst the blue and red horizontal branch 
components and the two observed subgiant branches also suggest that the RR Lyrae 
variables originated from the progeny of the bright subgiant branch. The RR Lyrae 
variables with a slightly enhanced helium ($Y$$\sim$0.270-0.280) have longer periods at a 
given amplitude, as is seen with Oosterhoff II (OoII) RR Lyrae variables, whereas the 
RR Lyrae variables with $Y$$\sim$0.248-0.270 have shorter periods, exhibiting properties 
of Oosterhoff I (OoI) variables. This correlation does suggest that the pulsational properties 
of RR Lyrae stars can be very useful for tracing the various sub-populations and can provide 
suitable constraints on the multiple population phenomenon.  It 
appears of great interest to explore whether this conclusion can be generalized 
to other globular clusters hosting multiple populations. 

 \end{abstract}
 

\section{Introduction}

Especially following the detection of two distinct subgiant branches (SGBs) in the 
color-magnitude diagram (CMD) of NGC$\,$1851 \citep{milone08}, attempts to piece together 
the formation history of this cluster have become alluring.  One promising explanation
of the split between 
the bright SGB (SGBb) and faint SGB (SGBf) is that the two subpopulations differ in age by 
about 1 Gyr, and this scenario has been discussed in a number of studies
\citep{milone08, carretta11a, carretta11b, gratton12}.  Another valid explanation is that
the SGB splitting is due to differing C+N+O contents and that the two SGBs are nearly coeval 
\citep{cassisi08, ventura09}.  The horizontal branch (HB) of NGC$\,$1851
is also bimodal, with both a prominent red HB clump and a blue tail. 
From the morphology of the HB and the main sequence (MS), strong helium variations within the 
cluster do not seem likely \citep{salaris08,dantona09}, and recent spectroscopy of the 
blue HB stars suggests minor
helium enhancements \citep{gratton12}.  Lastly, the red giant branch (RGB) is known to
harbor different populations \citep{grundahl99, calamida07, lee09, han09}.  

The stellar distribution along the HB of globular clusters is commonly used 
to understand their formation and evolution \citep{gratton10, dotter10}, and 
previous papers dealing with the modeling of the HB of 
NGC$\,$1851 \citep{salaris08, gratton12} used synthetic HB models to obtain scenarios of the formation 
this clusters bimodal horizontal branch.
In this paper we will approach a more detailed investigation of the portion of the
HB dealing with the instability strip (IS), by discussing the case of the RR Lyrae period distribution 
in NGC$\,$1851.

Period distributions of RR Lyrae stars have been shown to place strong constraints in the framework 
of canonical HB evolution.  For example, the problem of the peaked distribution of the RR Lyrae 
periods in M3 \citep{castellani81, rood89} has challenged model 
predictions as nicely described by \citet{catelan04}.
Recent studies have come to explain both the M3 period distribution and HB morphology as 
the consequence of a range of initial He together with a uniform total RGB mass loss (with a very small
spread) \citep{caloi08}, or a suitable bimodal mass-loss efficiency along the RGB but a single initial 
He abundance \citep{castellani05}.

The RR Lyrae properties of NGC$\,$1851 have been used to describe this cluster as 
``truly an unusual Oosterhoff type I object" \citep{downes04}.  This is largely because NGC$\,$1851 
is at the extreme end of the OoI-type clusters, with its RR Lyrae variables having not only a longer 
than average period for its Oosterhoff class, but also a ratio of first overtone RR Lyrae (RR1) to 
fundamental mode RR Lyrae (RR0) stars more in line 
with OoII-type GCs \citep{walker98}.  
The suggestion has also been made that the RR Lyrae variables can be divided into two subgroups 
based on their Ca $uvby$ photometry \citep{lee09}, although
they acknowledge that their sample is small, and the apparent bi-modality may merely reflect
a calcium metallicity spread in the variables.  
Further suggestions that 
this cluster may be different than other Galactic GCs comes from its phase-space distribution,
which indicates it may be associated with the Canis Major dwarf \citep[][however
see also L\'{o}pez-Corredoira et~al. 2007 who show that the signatures of the Canis
Major dwarf can be fully accounted for by Galactic models without new 
substructures]{frinchaboy04, martin04}.

Reproducing theoretically both the morphology of its unusual CMD as well as the period distribution 
of its RR Lyrae variables is an important step in piecing together the formation scenario
of NGC$\,$1851.

\section{RR Lyrae Observations }
\subsection{Sample and Completeness}
The most complete study of the RR Lyrae variables in NGC$\,$1851 was 
carried out by \citet{walker98}.  He presents 33 variables in a 13.6 arcmin$^2$ area centered 
on this cluster in the $B$,$V$ and $I$
passbands, 30 of which are RR Lyrae stars.  Recently \citet{sumerel04} 
discovered 19 additional variables and \citet{downes04} reported eleven variables, all within 
40" of the cluster center.  There is overlap in these two samples, as described by the
2011 update of NGC$\,$1851 in the \citet{clement01} catalog, and most of the new 
discoveries are RR Lyrae stars, although the 
classification for a handful of these stars is still uncertain.  Neither \citet{sumerel04} nor 
\citet{downes04} provide calibrated mean magnitudes or
amplitudes for their variables, making it difficult to use these stars in our analysis.
As there is no indication that the \citet{walker98} sample is incomplete at
distances greater than 40" from the cluster center,
we limit our sample of RR Lyrae stars with which to compare our HB models
to this outer region.
Of the 29 RR Lyrae stars studied by \citet{walker98} that are greater than 40" from the 
center, 25 have both unblended magnitudes and well determined periods.  
Thus our sample of variables is 25/29 or 86\% complete (outside the inner core).

\begin{figure}[htb]
\includegraphics[width=0.95\hsize]{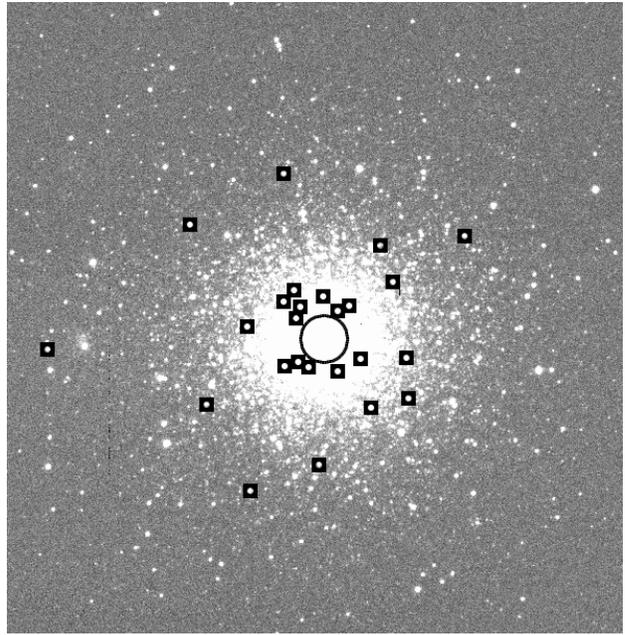}
\caption{Image of NGC$\,$1851 showing the 25 RR Lyrae variables used in our analysis.
The central 40" is designated by a circle.  This region is known to harbor RR Lyrae variables, 
but due to the severe crowding, no calibrated photometry or amplitudes exist for the RR Lyrae stars
in this region.
\label{1851imga}}
\end{figure}

The position on the sky of our sample of variables is shown in Figure~\ref{1851imga}, and 
the central 40" is designated by a circle.  
Light curves for these stars are presented by \citet{walker98} in the $BVI$ passbands, so for these 25 
RR Lyrae variables, robust mean magnitudes, periods and amplitudes are available.
The edges of the instability strip, judged by the measured colors
of variables near the strip boundaries, were also determined by \citet{walker98}, as 
well as the RR1 - RR0 boundary.

\subsection{Period-Amplitude Diagram}
\begin{figure}[htb]
\includegraphics[width=1\hsize]{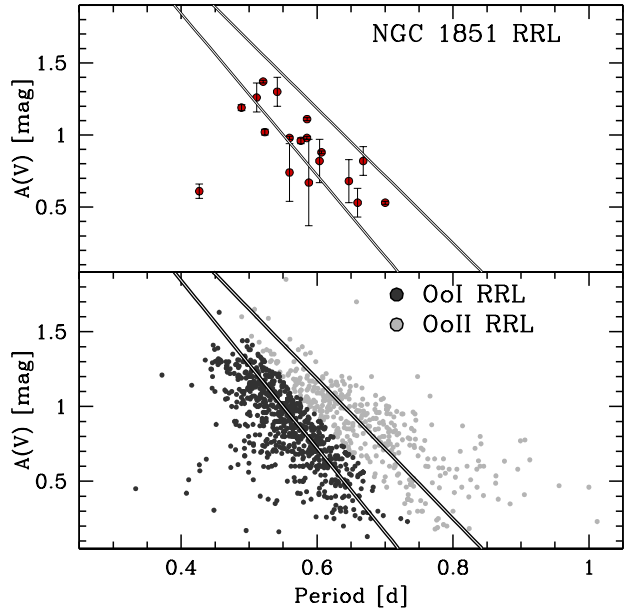}
\caption{ {\it Top:} Period-amplitude diagram for our sample of RR0 Lyrae variables in NGC$\,$1851.
{\it Bottom:} Period-amplitude relation for 1097 RR0 Lyrae variables in 39 Galactic GCs.
Our division of OoI and OoII-type RR Lyrae variables are shown
by dark and light circles, respectively.
The lines derived by \citet{clement99a} for Oosterhoff I and Oosterhoff II RR0 are overplotted.  
\label{PAall}}
\end{figure}

\citet{clement99a} show that the position of an RR Lyrae variable in the 
period-amplitude ($\rm PA$) diagram is not a function of metal abundance, 
but rather of Oosterhoff type, and derive 
PA relations for OoI and OoII-type RR Lyrae stars empirically.  More recently
\citet{cacciari05} study the $\rm PA$ plane 
of 3 typical OoI-type GCs, 3 typical OoII-type GCs and 3 intermediate types and find that 
there is a unique period-amplitude relation independent of metallicity for RR0 variables 
in OoI-, OoII- and intermediate-type clusters.  
The periods and $V$-amplitude of our sample of RR0 Lyrae variables is shown in
Figure~\ref{PAall}, and the period-amplitude relation of typical
OoI and OoII-type systems is over-plotted.
We note that although many of the RR Lyrae stars in NGC$\,$1851 have periods and 
amplitudes that cause them to fall near the OoI PA relation, there 
are a number of stars following the OoII PA relation.

It is well known that the Blazhko effect, or other
effects such as a rapidly changing period, can cause scatter in the
PA plane \citep{clement99a}.  The Blazhko effect causes the amplitude of light variation 
to vary over timescales longer than the basic pulsation period. 
The \citet{walker98} RR Lyrae variables were observed over an ample time frame (126 total frames
observed over 15 nights during a 1.5 year time span), so
determining the amplitudes using the average light curves of the RR Lyrae stars is
straightforward.
Nevertheless a visual determination of the change in amplitudes in each RR0 is 
obtained and shown as an error-bar in Figure~\ref{PAall}.  The change from the average light curve
amplitude to the maximum Blazhko amplitude ranges from $\sim$0.1 - 0.3 mag, and photometric
uncertainties lead to amplitude uncertainties of $\sim$0.01-0.05 mag.  
We conclude that even when taking amplitude variations into account, the variables are 
both OoI and OoII-type.

In comparison, Figure~\ref{PAall} shows the periods and $V$-amplitudes of 1097 RR0 
Lyrae variables in 39 Galactic globular clusters.  The data for this diagram come from 
Table~\ref{GClist}, where each cluster is listed along with the number of RR0 Lyrae stars 
within the cluster that have well determined periods and $V$-amplitudes.  For completeness, the $\rm [Fe/H]$ from \citet{carretta09}, HB-type and Oosterhoff type for each GC are also given.
This sample of RR0 Lyrae variables are divided by their position in the period-amplitude plane following
the lines that \citet{clement99a} derived for Oosterhoff I and Oosterhoff II RR0 stars (see
Figure~\ref{PAall}).  
Here an OoI RR Lyrae variable is defined by
\begin{equation}
A(V) > -5.1453 \ P + 4.02
\end{equation}
and an OoII-type RR Lyrae variable by
\begin{equation}
A(V) < -5.1453 \ P + 4.02,
\end{equation}

\noindent
where $A(V)$ is the $V$-amplitude and $P$ is the period.
We define an Oosterhoff ratio for each GC, which is simply the number of OoI-type RR0 Lyrae stars
compared to the total number of RR0 Lyrae stars in the GC, $\rm OoI_{RR0} / Tot_{RR0}$.

We note that \citet{cacciari05} showed for the M3 RR0 Lyrae stars, a quadratic PA relation 
is a closer fit than a linear one.  However, their relation 
does not approximate high amplitude RR0 Lyrae variables well
($\rm A_V$ $>$ 1.5 mag), largely because such variables are absent in M3.  
As our comprehensive sample includes a handful of such stars (one OoI- and three 
OoII-type RR Lyrae), the \citet{cacciari05} relation is not used here.

\begin{figure}[htb]
\includegraphics[width=1\hsize]{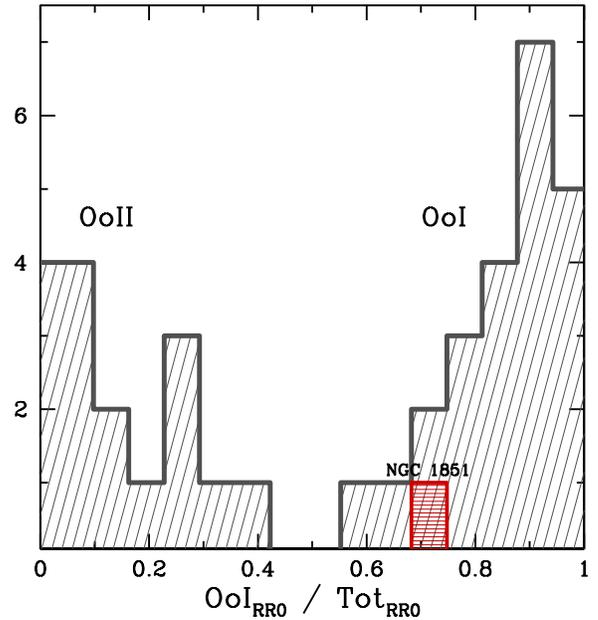}
\caption{A histogram of the ratio of OoI stars in the Milky Way GCs.  The Oosterhoff ratio of the
NGC$\,$1851 RR0 Lyrae stars is highlighted.  
\label{ooratio}}
\end{figure}

Figure~\ref{ooratio} shows the histogram of the Oosterhoff ratio of the GCs in our sample.
Most of these clusters lack a complete sample of RR Lyrae variables and at least some of
the RR Lyrae amplitudes are likely affected by the Blazhko effect or other
light curve ``noise".  But even with these caveats, it is clear from the figure 
that the Oosterhoff ratio splits the GCs into two groups; the 
OoI-type clusters have RR Lyrae variables with shorter periods for a given 
amplitude and hence have larger Oosterhoff ratios (with respect to the OoII-type
clusters).  Further, there is an absence of clusters 
falling in the ``gap".  We therefore believe that our Oosterhoff ratio is useful
to distinguish between OoI- and OoII-type GCs.  Moreover, this ratio may be used
to evaluate the degree for which a GC is a typical OoI- or OoII-type cluster.  

The majority of OoI-type GCs have a RR Lyrae star population in which 
$\rm OoI_{RR0} / Tot_{RR0}$ $>$ 0.8 (i.e., 80\% of the
variables can be defined by Equation~1).  
In contrast, the Oosterhoff ratio of NGC$\,$1851 is 0.74.  Equation~1 and 2 are defined
somewhat arbitrarily, and therefore we experimented with a variety cuts (where we
modified the zero-points and slopes) to distinguish between OoI- and
OoII-type stars.  The Oosterhoff ratio of NGC$\,$1851 varied between $\sim$0.63-0.75,
a percentage that is smaller than 83\% of the rest of the OoI-type clusters.
This indicates that NGC$\,$1851 contains variables with more OoII-like periods and
$V$-amplitudes than the majority of the other Milky Way OoI-type GCs.  

The other three OoI GCs that have comparatively small Oosterhoff ratios
are NGC$\,$362 (0.71), NGC$\,$4590 (0.58) and NGC$\,$6362 (0.67).  
It has been noted that the RR Lyrae variables of NGC$\,$362 and 
NGC$\,$1851 are remarkably similar in the period-amplitude diagram, suggesting similar 
masses and luminosities.  This is also seen here, as they have very similar Oosterhoff ratios.
Recently it was found that like NGC$\,$1851, NGC$\,$362 also has a split sub-giant branch, 
although the SGBf component includes only a few percent of the total number of SGB stars of the cluster 
(just $\sim$2-3\%, A. Milone private communication).   

New photometry of the NGC$\,$1851 RR Lyrae variables, in which periods and
amplitudes of the inner RR Lyrae stars have been obtained,
suggest that NGC$\,$1851 has a slightly higher Oosterhoff ratio of $\sim$0.80
(Amigo et~al 2012, in preparation).  We therefore suspect that the Oosterhoff 
ratio of NGC$\,$1851, whereas still lower than the majority of the OoI-type
clusters, is not as abnormal as the globular clusters 
NGC$\,$362, NGC$\,$4590 and NGC$\,$6362,
discussed above.  

\subsection{First Overtone vs. The Fundamental Mode}
The number ratio of first-overtone RR Lyrae stars to total RR Lyrae stars, $N_{\rm 1}/N_{\rm tot}$, 
is usually given to quantify the frequency of
the different RR Lyrae pulsators.  In general OoII-type GCs have about 2-4 times 
the frequency of RR1 stars as OoI-type GCs.  This is thought to be related to the
transition temperature between the instability strip for first overtone
pulsation and fundamental one.  For example, moving the transition from 
RR0 to RR1 variables toward lower temperatures (i.e., transforming fundamental 
in first overtone pulsators) has the twofold effect of increasing the periods of the 
RR0 Lyrae population, as well as increasing the relative number of first overtones.
This is discussed in detail by Castellani, Caputo \& Castellani (2003).

NGC$\,$1851 has a large $N_{\rm 1}/N_{\rm tot}$ with respect to most OoI-type GCs.  
This ratio strongly depends on the completeness of the sample, and 
because NGC$\,$1851 is too crowded for ground-based observations to resolve, 
the \citet{walker98} sample is incomplete at distances close to the core.  
Assuming the \citet{walker98} RR Lyrae sample is complete at distances larger than
40" (see Figure~\ref{1851imga}), $N_{\rm 1}/N_{\rm tot}$ = 0.27, one of the largest for
OoI-type GCs.  For example, from the \citet{castellani03} compilation of 32 clusters with 12 
or more pulsators and well recognized period and pulsation modes, only 3 of the 17 OoI GCs have an
$N_{\rm 1}/N_{\rm tot}$ greater than 0.27.  
Recent studies have discovered RR Lyrae stars closer than 40" from the center.
\citet{downes04} find $N_{\rm 1}/N_{\rm tot}$ = 0.54, and using Clement's catalog, 
the value of this ratio is $\sim$0.40.  These ratios are somewhat uncertain, however, 
because of the lack of mode identification for some of these newly discovered 
pulsators.  The periods
listed in \citet{sumerel04} are ``tentative" and the periods derived by
\citet{downes04} do not cover a full period for any of their objects.
Further, because these stars lack amplitudes, their position in a PA diagram
also can not be used as a diagnostic to identify fundamentals and first overtones. 
Although their is no consensus on the value of $N_{\rm 1}/N_{\rm tot}$ yet, it is 
clear that this ratio is at least 0.27, and likely even larger.  A $N_{\rm 1}/N_{\rm tot}$ 
larger than 0.27 is also consistent with the results from Amigo et al. (2012, in preparation).

\section{Synthetic HB Modeling}
Previous synthetic HB models for NGC$\,$1851 have been
presented by \citet{catelan98}, and more recently by \citet{salaris08}, \citet{han09} and 
\citet{gratton12}.  \citet{salaris08} compared their simulations to the $HST$ observations by 
\citet{milone08} and found 
two satisfactory scenarios to reproduce the CMD of HB stars. 
In both of these models, the blue HB, red HB and variable stars are predicted to come from 
the SGBf, and the stars from the SGBb are confined to the red portion of the observed sequence. 
This inference was based on the number ratio SGBf/SGBb=55:45 determined 
in \citet{milone08}. \citet{salaris08} also found that the initial He abundance of HB stars had to be 
relatively uniform to reproduce the CMD derived by \citet{milone08}. 
In these data the measured magnitudes and colors of the RR Lyrae population are at random phases; 
therefore the portion of the observed HB crossing the IS could not be used for
detailed constraints on the models.  However, they did verify that recent theoretical 
pulsation models of RR Lyrae stars \citep{dicriscienzo04} predict an 
instability strip for NGC$\,$1851 too red by $\sim$0.03$-$0.04 mag in ($F606W$-$F814W$) 
compared to the \citet{milone08} data. 

\citet{han09}, on the other hand, found from their $UVI$ photometry that the RR Lyrae 
variables could come from both the
SGBb and the SGBf.  They construct two population models for NGC$\,$1851.
In the first, the second 
generation population is more enhanced in metallicity but not in helium ($\Delta$Z=0.0004 and 
$\Delta$age = 0.1Gyr), and in the second, both metal and helium abundances are 
enhanced ($\Delta$Z=0.0004, $\Delta$Y=0.05, and $\Delta$age = 0.1Gyr).  They find that their
$\Delta$Z-only model is in conflict with the observed CMDs of NGC$\,$1851, but that their 
$\Delta$Z+$\Delta$Y model is in good agreement with the observations from the MS to the HB.
In this model, the RR Lyrae variables exhibit different He abundances.  Because
their RR Lyrae variable sample was found at a random phase of pulsation, the RR Lyrae
colors could not be used as a stringent constraint in their models and the RR Lyrae star
periods were not discussed.

More recently \citet{gratton12} considered the revised SGBb/SGBf ratio determined 
by \citet{milone09} and introduced new spectroscopic constraints;
they find a small difference in the iron content between the SGBb
and SGBf, and argue that an age spread of $\sim$1.5 Gyr is the most viable explanation for the 
splitting SGB.  They also find that the RHB stars separate into two groups depending on
their O and Na abundances, and that the BHB stars are slightly helium enriched as compared
to the RHB stars.  Hence to satisfy these constraints, the HB is modeled with four different 
components, with the IS originating from the SGBb.  That each SGB hosts multiple generations
of stars is shown from spectroscopy of stars on the double SGB \citep{lardo12}, making it
likely that multiple components may be needed to model the HB. 

Our own synthetic HB calculations described below are aimed at answering the following 
question: what is the most straight-forward way to reproduce the RR Lyrae instability strip of
NGC$\,$1851 -- and in particular the pulsational properties 
of its RR Lyrae variables?  We seek to provide
a simple and attractive explanation for the cluster HB and IS morphology, 
keeping the number of free parameters to a minimum, yet still reproducing the RR Lyrae
star properties that make this cluster stand out as having an unusual Oosterhoff type.

\subsection{Synthetic HB models}
The HB evolutionary tracks used here are from the BaSTI stellar 
library \citep{pietrinferni04, pietrinferni06, pietrinferni09} and have already been described 
by \citet{salaris08} and \citet{cassisi08}.  They are also the same that \citet{gratton12} employed.  
Briefly, evolutionary tracks are for a normal $\alpha$-enhanced ([$\alpha$/Fe]=0.4) 
metal mixture, with $\rm [Fe/H]$=$-$1.31 dex and $Y$=0.248.   
The HB tracks were interpolated among the models with $Y$=0.248 and additional BaSTI models with 
$Y$=0.300, to determine HB tracks for intermediate values of $Y$,
at the same iron content.  Similarly, to determine HB tracks with a milder CNO-enhancement,
an interpolation between the reference set and the models 
with the CNO sum enhanced by 0.3 dex \citep{pietrinferni09} is used
for a portion of the synthesis, as in \citet{gratton12}.
Hence, the $\rm [Fe/H]$, $\alpha$-enhancement and CNO-enhancement is 
consistent with spectroscopic results from \citet{carretta11a} and \citet{gratton12}. 
We wish to remind the reader that, as long as the CNO sum is unchanged, the effect of the observed CNONa 
anticorrelations (overimposed to a standard $\alpha$-enhanced metal mixture) 
on the evolutionary tracks and isochrones is negligible, and standard $\alpha$-enhanced models 
are adequate to represent the whole cluster population. Only an enhancement of the CNO sum 
requires the calculation of appropriate models. On the other hand CNONa anticorrelations even at constant CNO 
affect the bolometric corrections of filters like $B$ and $U$ \citep{sbordone11} at low effective 
temperatures, but not longer wavelength filters.

The four HB components described by \citet{gratton12} are used as a starting
points for our calculations. 
Objects from our synthetic HB that fall within the observed IS from \citet{walker98} are
considered RR Lyrae variables (this region is labeled in Figure~\ref{cmd_hb}) and their period is 
calculated from the pulsation equation given by \cite{dicriscienzo04}.  The intensity
mean magnitudes and colors given by \citet{walker98} are used as a comparison to the synthetic
HB, because the static magnitudes and colors from stellar evolution models are represented
better by intensity-averaged quantities \citep{dicriscienzo04}.
Although the \citet{walker98} observations include the $BVI$ passbands, we 
employ only the $V$ and $I$ magnitudes, because -- as discussed before -- 
they are not affected by the observed CNONa abundance anticorrelations.

In addition to the observed $V$ and $(V-I)$ distribution of the HB stars in the CMD, and the observed 
$\rm (B\thinspace{:}V\thinspace{:}R)$ (blue\thinspace{:}variable\thinspace{:}red HB) 
ratio of $\rm (B\thinspace{:}V\thinspace{:}R)$ = $\rm (33\pm8\thinspace{:}10\pm5\thinspace{:}56\pm11)$
\citep[in line with the results by][]{catelan98, saviane98},   
we impose as a further constraint on our simulation the observed 
distribution of the RR Lyrae periods.

As in \citet{gratton12} we adopted $\rm E(B-V)$=0.02~mag \citep{walker98} 
and fixed the apparent distance modulus to $\rm (m - M)_V$=15.56~mag 
by matching the observed mean magnitude of the RHB with our synthetic counterpart. 
The $\rm (B\thinspace{:}V\thinspace{:}R)$ ratio of 
our `best fit' simulation is $\rm (27\thinspace{:}9\thinspace{:}64)$, consistent with the observed 
value.  For reasons that will become clear in the discussion that 
follows, we consider a preliminary reference age of 11~Gyr for the progenitors of the RHB stars. 
This implies, for the assumed metallicity and a `normal' $Y$=0.248, 
an initial mass of 0.86${\rm M_{\sun}}$ for the stars at the tip of the RGB.  The 
HB components are described below. 

(1) As in \citet{gratton12}, the majority of the RHB population is modeled
with normal CNO abundance, a normal $Y$=0.248, and a Gaussian mass distribution with 
${\rm <M>}$ = $\rm 0.67\pm0.005~M_{\sun}$. This corresponds to a total mean mass loss 
$\Delta {\rm M}$=0.19~${\rm M_{\sun}}$ along the RGB, for the assumed 11~Gyr age.

(2) A smaller RHB subpopulation, that is Ba-rich, 
makes up $\sim$10\% of the HB population. It is modelled, as in \citet{gratton12}, 
employing a Gaussian mass distribution with $<M>$ = $\rm 0.65\pm0.004~M_{\sun}$ 
(corresponding to $\Delta {\rm M}\sim$0.21~${\rm M_{\sun}}$) an 0.15~dex 
enhanced CNO abundance, and normal $Y$=0.248. 
If we assume that the mean total mass loss has to be constant among all cluster RGB stars 
-- and equal to $\Delta {\rm M}$=0.19~${\rm M_{\sun}}$ as determined for the rest of the  
RHB component -- the mean value of the mass for this HB sub-population implies an age $\sim$1~Gyr older 
for the progenitors of this HB component.

(3) The horizontal part of the BHB, including the RR Lyrae instability strip, 
makes up $\sim$10\% of the cluster
stellar content.  This component is the focus here, and is the only one that is modified
from \citet{gratton12}.  In particular, instead of adopting a 
constant He abundance $Y$=0.265, the helium
content for stars between the blue end of the red clump and the beginning of the BHB blue tail 
has a continuous distribution between $Y$=0.248 and 0.280. A simple, flat
probability distribution for $Y$ and constant 
$\Delta {\rm M}$=0.19~${\rm M_{\sun}}$ (for an age of 11~Gyr) 
with a 1$\sigma$ Gaussian spread of 0.005~${\rm M_{\sun}}$ -- as for the RHB stars -- 
for all RGB progenitors provide a good match to the observed RR Lyrae 
periods, as discussed below. 

The mean He abundance in the IS is $<Y>$=0.271, close to 
the constant abundance $Y$=0.265 employed by \citet{gratton12} for this component,
and the mean mass is ${\rm <M>}$ = $\rm 0.634~M_{\sun}$. 
It is worth noting that the observed HB distribution of these stars is well matched by both
this simulation and the simulation by \citet{gratton12}.  This spread in He content is 
necessary to reproduce the observed period distribution.

(4) The blue tail of the BHB population makes up $\sim$20\% of the HB stars. 
As in \citet{gratton12} this component is modeled with normal CNO abundance 
and $Y$=0.28. The Gaussian mass distribution has ${\rm <M>}$=$\rm 0.59\pm0.005~M_{\sun}$, 
that for an age of 11~Gyr would correspond to a mean $\rm \Delta M$=$\rm0.22~M_{\sun}$. 
If the mean total mass loss is instead fixed at $\Delta {\rm M}$=0.19~${\rm M_{\sun}}$, this 
value of ${\rm <M>}$ implies that the progenitors of the BHB blue tail stars are $\sim$1.5~Gyr 
older that the RHB 
component with normal CNO and $Y$.  Notice that the constraint on the progenitor $Y$ is 
weaker for BHB stars \citep[see discussion in][]{gratton12} and a small spread of order 0.01 
may be present.

These results from the HB synthetic modeling can be interpreted in terms of the progeny of the 
SGBb and SGBf subpopulations (the ratios $\sim$ 2/3 and $\sim$ 1/3 of the total SGB population, 
respectively, are adopted as determined by Milone et al. 2009) as follows:
\\
(1) The sum of the fraction of stars in the blue tail of the BHB and in the mildly CNO-enhanced 
Ba-rich RHB component is $\sim$35\% of the total HB population.  If we consider as a 
reference the CMD of 11~Gyr old SGB stars with `normal' $Y$ and CNO abundance, the 
progenitors of these two HB components will be distributed along a fainter SGB than the 
reference one. In the case of the progenitors of the BHB component this is an age effect, for 
a change of $Y$ does not have a major effect on the SGB luminosity. For the Ba-rich RHB 
progenitors the reason is the slightly higher age and the mildly enhanced CNO abundance, 
that act both in the direction of producing a fainter SGB.  As a result, both the 
Ba-rich RHB stars and the blue tail HB progenitors display an approximately 
coincident SGB, that we tentatively identify as the SGBf in the cluster CMD.
\\
(2) The sum of the fraction of stars in the horizontal BHB (including the IS) and the RHB stars 
with normal composition amounts to $\sim$65\% of the total HB population. 
We identify their progenitors as the stars harbored by the SGBb in the cluster CMD.

To conclude this section, it is worth noting that the exact value of the assumed 
reference age (11~Gyr) is not critical.  Had a different age been assumed, i.e. 10 or 12~Gyr, the 
previous conclusions will still be valid.  The only difference is that all values of $\Delta {\rm M}$ 
would need to be shifted downwards (or upwards) by $\sim$0.02$~{\rm M_{\sun}}$-- to keep the 
mass distribution along the HB unchanged -- but the 
interpretation of the results would be identical.  Finally, the different chemical composition -- and 
small age differences -- assigned to the SGBb and SGBf populations  
do not affect substantially SGB and RGB timescales; as a consequence, 
the number ratio SGBb/SGBf will be approximately equal to the number ratio of their HB progeny.

\subsection{Comparison With Observations}

\begin{figure*}
\includegraphics[width=0.9\hsize]{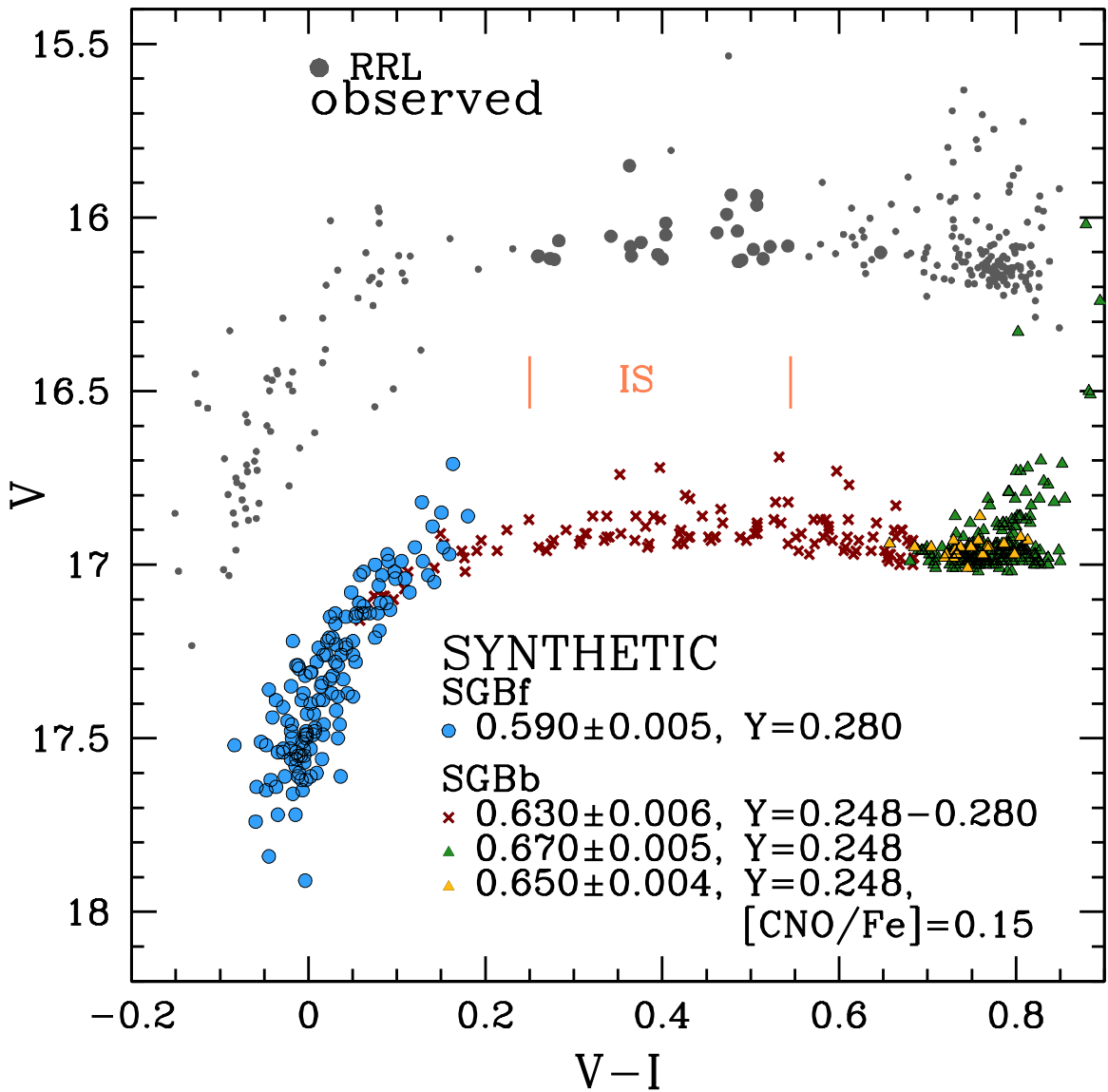}
\caption{Qualitative comparison of the observed HB and the synthetic HB, assuming
four separate HB components.
The RR Lyrae instability strip is marked, and corresponds to that found by the
\citet{walker98} study of RR Lyrae variables in NGC$\,$1851.  The \citet{walker98}
RR Lyrae variables are designated by large circles.
\label{cmd_hb}}
\end{figure*}

This synthetic HB model is shown compared to the observed one in 
Figure~\ref{cmd_hb} where an observational scatter of $\sigma_{V,I} =$ 0.01~mag is assumed.  
The four components are high-lighted for clarity and the RR Lyrae region is labeled.
Our full synthetic HB model reproduces two peculiarities in the CMD of NGC$\,$1851
pointed out by \citet{brocato99}, namely
the clump of stars near the red edge of the HB and the slightly tilted 
HB ($\rm \Delta_V^{tilt}$$\sim$0.1 mag).  Features such as these are present also in NGC$\,$6362 (which 
has an RR Lyrae population with properties very similar to
NGC$\,$1851), and in the extreme cases of NGC$\,$6388 and NGC$\,$6441 
($\rm \Delta_V^{tilt}$$\sim$0.5 mag).  

The focus here concerns the component that includes the instability strip, which comes
from the SGBb.  
As \citet{walker98} mentioned, the ZAHB is very cleanly defined and is not horizontal,
being slightly brighter at bluer colors.  This is reproduced in our synthetic HB model by
stars that range in helium abundance from $Y$=0.25-0.28, and
range in mass from 0.61 to 0.65$\rm M_{\odot}$.  

\begin{figure}[htb]
\includegraphics[width=1\hsize]{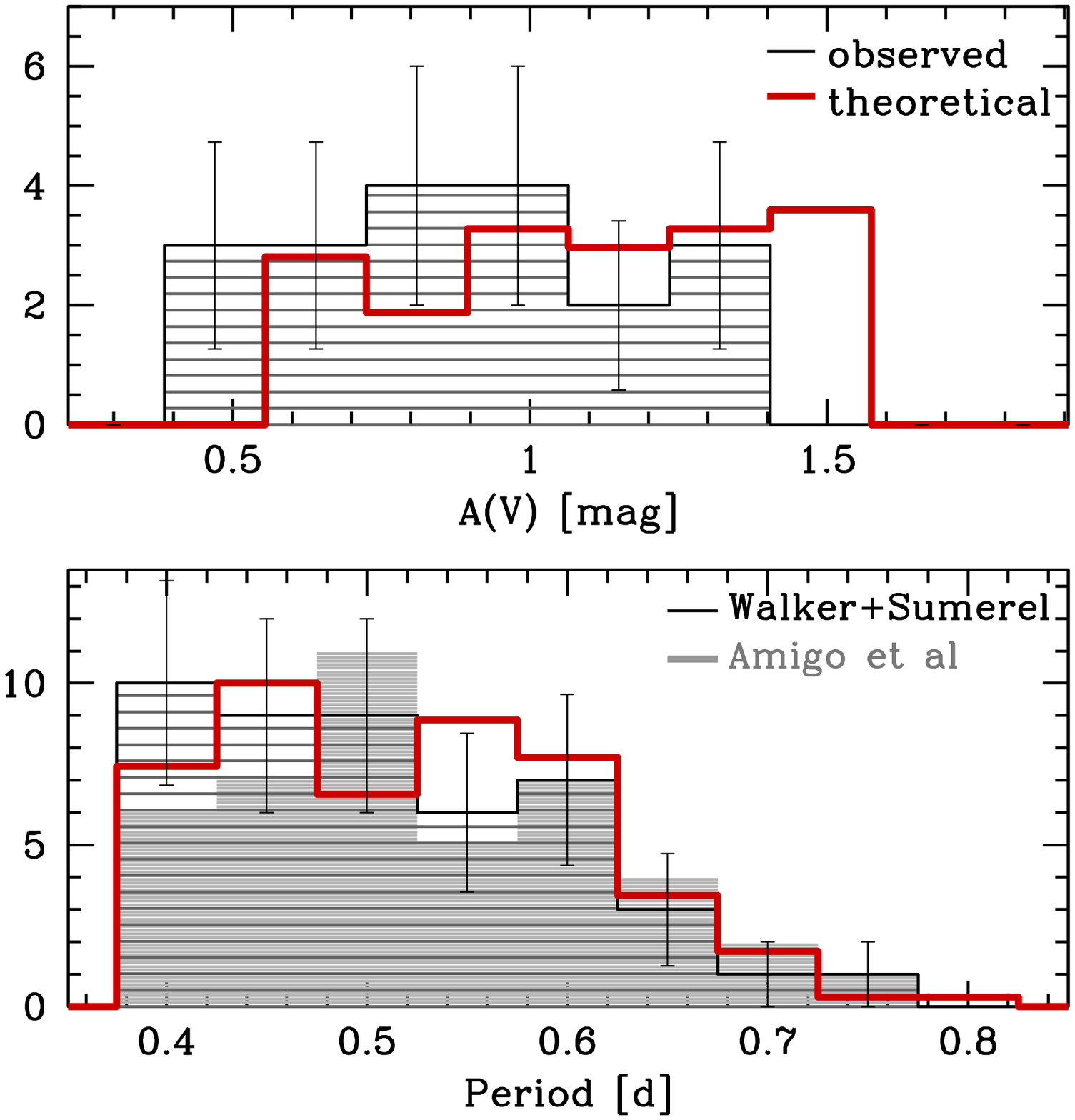}
\caption{A comparison between the theoretical periods and amplitudes from our
synthetic HB versus the observed periods and amplitudes.  A KS test indicates that 
statistical significance is detected between the observed and 
theoretical periods and amplitudes.
\label{PAhist}}
\end{figure}

The theoretical periods and pulsation amplitudes from the RR Lyrae variables
in our synthetic HB model are compared to the observed periods and amplitudes
in Figure~\ref{PAhist}.  
Here the observed periods come from 28 RR0 and 18 RR1 variables as determined by 
\citet{walker98} and \citet{sumerel04} to encompass all the data available in the
literature \citep[see the 2011 update of NGC$\,$1851 in][]{clement01}.  
We also compare the theoretical periods with 27 RR0 and 18 RR1 variables 
as determined by Amigo et~al. (2012, in preparation).  These authors derive periods for
the recently identified inner RR Lyrae variables based on
light curves with $\sim$200-300 points in each of the $B$- $V$- and $I$-passbands.
The observed amplitudes come exclusively from the \citet{walker98} RR0 Lyrae 
sample, as \citet{sumerel04} present instrumental magnitudes only. 
The amplitudes from the Amigo et~al. (2012, in preparation) sample are not used, because unlike 
when determining periods, amplitudes can be affected by
crowding and blending issues \citep[e.g.,][]{majaess12} and we do not have a feel for how/if blending
affects their (preliminary) amplitude determinations.
  
\citet{marconi11} provided a detailed comparison between the impact of the He abundance 
on the pulsation properties of RR Lyrae stars and concluded that He content marginally 
affects the pulsation behavior of RR Lyrae stars.  They noted that the increase in the average 
pulsational period associated with the He increase is only due to the brighter luminosities which 
characterize He-enhanced evolutionary models. As a consequence, from a theoretical point
of view the impact of an He-enhancement on the pulsation properties of RR Lyrae stars can be 
directly taken into account when adopting evolutionary tracks for the appropriate He abundance and 
pulsational model predictions obtained for a canonical He abundance.

The periods of the RR1 variables are fundamentalized via 
log $\rm P_0 \sim log P + 0.127$, where $\rm P_0$ is the fundamental mode period, and 
the theoretical periods are calculated for all HB stars falling within the observed IS using the
\citet{dicriscienzo04} RR Lyrae pulsation models.
The mean observed RR0 Lyrae period in NGC$\,$1851 is $\rm <P_0>$ = 0.571 \citep{catelan09}, 
and from our model $\rm <P_0>$ = 0.569 $\pm$ 0.006 d is calculated, where the 
uncertainty is the error in the mean.  The number of RR Lyrae stars
in the simulations is $\sim$5 times larger than the number of observed
RR Lyrae stars.  In this way, in the synthetic HB model, the effect of statistical fluctuations in the
number of objects at a given magnitude and color is minimized.  

The observed minimum fundamentalized period of the first overtones 
and the minimum fundamentalized period from
the synthetic HB is very similar, $P_F$$\sim$0.34 days.  It has been shown that the minimum 
fundamentalized period is a robust observable to 
constrain the evolutionary properties of RR Lyrae stars \citep{bono95}, so it is especially
encouraging that the observed and theoretical values agree.

We have performed a Kolmogorov-Smirnov (KS) test to establish whether one can reject 
the null hypothesis that the two samples of periods (observed and synthetic) come from the same 
distribution.  From a comparison between the combined \citet{walker98} and \citet{sumerel04} sample
and the synthetic RR Lyrae stars,
the KS-test returns a probability P=0.86, well above the default threshold 
${\rm P_{th}}$=0.05 below which one rejects the null hypothesis.
When the Amigo et al. period distribution is compared with our 
theoretical one, if we assume that both samples are drawn from the same parent 
population, the samples will differ by this amount or more 94\% of the time.
On this basis, we find that the synthetic
periods from our simulated HB and the observed periods agree well 
with each other. 

Theoretical pulsation amplitudes are also determined from the 
\citet{dicriscienzo04} relations, although three points should be taken into account:

1.  Theoretical amplitudes depend on the the mixing-length, $l/H_p$, 
(where $l/H_p$ is the pressure scale-height), which is uncertain 
and may change from the blue to the red edge of the IS \citep{marconi03}.
Pulsation amplitudes
are affected by $l/H_p$ in the sense that a larger value of $l/H_p$ translates to 
smaller pulsation amplitudes as a consequence of the larger efficiency of convective 
motions and, hence larger quenching to the pulsation mechanism provided by convection.

2.  The synthetic pulsational amplitude  - period predictions provided by \citet{dicriscienzo04} 
show a large scatter, of the order of $\sim$0.2 mag (standard deviation) in the $V$-amplitude 
at a fixed period (see Fig. 6 of their paper, where the dashed lines represent the standard deviation). 
Since our predictions of the $V$-amplitude in our synthetic HBs are based 
on an $A(V)$ - period relation obtained by an interpolation on the data provided by 
\citet{dicriscienzo04}, we do not expect a great accuracy in our $A(V)$ estimates. 
In addition, \citet{dicriscienzo04} have shown that for $P<0.68$ d, pulsation model 
predictions for the $V$-amplitude are still more affected by a change in the 
adopted mixing length value (see lower panel of the quoted figure): for increasing 
mixing length values they predict a significant decrease in the dependence of
the $V$-amplitude on the pulsation period.

3.  When dealing with pulsating structures, the static magnitudes differ from the
measured magnitudes, which are mean quantities averaged over the pulsation
cycle.  In finding theoretical amplitudes, a correction between static and intensity-averaged magnitudes 
is required.  The discrepancy between static and mean values is a function
of the pulsation amplitude, and the corrections adopted here come from
\citet{bono95}.

When using a $l/H_p$=2.0, the \citet{walker98} $V$-amplitude distribution is similar to that 
observed, although the theoretical amplitudes appear to be $\sim$0.1 mag larger.  
This is not completely surprising considering the scatter in the \citet{dicriscienzo04} $V$-amplitude relation 
as well as the magnitude corrections discussed above.  
Extrapolating linearly between 
$l/H_p$=1.5 and 2.0, an increase in the mixing length of 0.1 would cause a decrease of the
theoretical amplitudes by $\sim$0.08 mag. Such a decrease would provide a satisfactory 
agreement between theory and observations. A  0.1 change in $l/H_p$ is well within
the uncertainties in the mixing length calibration, and not nearly large enough to
affect the predicted pulsation  periods \citep{bono94, dicriscienzo04b, marconi07}.
An $l/H_p$=2.0 was also used by \citet{bono07} to derive a visual distance modulus from 
nonlinear convective models of RR Lyrae stars to NGC$\,$1851 of 
$(m-M)_V$=15.52$\pm$0.11 mag, which is similar to the distance modulus adopted here.
In contrast, when using a $l/H_p$=1.5, \citet{bono07} 
find a distance modulus of $(m-M)_V$=15.40$\pm$0.12.  
Therefore we conclude that 
using a larger value of $l/H_p$, ($l/H_p$$\sim$2.0), provides a consistent comparison 
between our synthetic HB and pulsational predictions.

In general, the RR Lyrae variables with $Y$ $<$ 0.27 fall in the OoI area of the PA diagram, 
whereas the RR Lyrae variables with $Y$ $>$ 0.27 fall close to the OoII line.  
Assuming that the period-amplitude diagram can be effectively used to classify RR Lyrae 
stars into an Oosterhoff type, this means that He and Oosterhoff type are correlated
in this cluster.  This is not completely unexpected, as an increase in He makes RR Lyrae 
variables brighter and, as a consequence, higher helium abundance makes their 
pulsational period longer \citep{bono97, marconi09}.

On a general ground, for a given total mass the HB stars with $Y$$<$0.27 are redder 
than those with $Y$$>$0.27.  The red part of the IS, where the fundamental mode
RR Lyrae stars reside, is consequently more populated, and a smaller $N_{\rm 1}/N_{\rm tot}$
is obtained.  Our synthetic HB yields $N_{\rm 1}/N_{\rm tot}$ $\sim$ 0.1 for the stars with $Y<$0.27.
This is a ratio that is seen for the majority of the OoI-type GCs.
In contrast, the blue part of the horizontal branch contains more stars with higher helium
abundances, and as the first overtone RR Lyrae variables reside in the blue part of the IS, a larger
$N_{\rm 1}/N_{\rm tot}$ is obtained.  Our synthetic HB yields $N_{\rm 1}/N_{\rm tot}$$\sim$0.45 for 
stars with $Y$$>$0.27, a ratio more in line with OoII-type GCs.  The observed ratio 
of first overtone to total RR Lyrae variables for NGC$\,$1851 in our sample is 
$N_{\rm 1}/N_{\rm tot}$ $\sim$0.30 and is easily explained (and reproduced
with our synthetic HB) by the spread in $Y$ along the RR Lyrae instability strip.  

We find that simulations using a constant helium for the portion of the HB containing 
the IS \citep[as in][]{gratton12} do not fit the constraints given by the NGC 1851 RR Lyrae
variables as well. For example, adopting $Y$=0.265 results in an $\rm N_1/N_{tot}$ = 0.11 
(versus the observed $\rm N_1/N_{tot}$ = 0.28).

We note that \citet{milone08} provide an upper
 limit to a possible dispersion in helium abundance 
of $\rm \Delta Y$=0.026  between the two SGBs in NGC$\,$1851, a value close to  
the spread assumed in the synthetic HB presented here.  Other estimates of the He spread in
NGC$\,$1851 are slightly larger, i.e., $\rm \Delta Y$=0.04 \citep{ventura09},
$\rm \Delta Y$=0.05 \citep{han09} or $\rm \Delta Y$=0.048 \citep{gratton10}.

\section{Discussion and Conclusions}
The population distribution of the stars along the HB 
has been modeled assuming the presence of (at least) four populations with differing helium contents.
In our simulations, the only parameter we vary is the initial He abundance of the HB 
progenitors, keeping the same total RGB mass loss for all components.
Both the RR Lyrae period distribution as well as the number ratio of first overtone RR Lyrae to total RR Lyrae
stars, $\rm N_1/N_{tot}$, provides constraints pertaining to the component of the HB containing the IS.
It is straight forward to reproduce the observed distribution of RR Lyrae stars inside the instability
strip with minor He variations ($Y$$\sim$0.248-0.280)
and from a HB subpopulation corresponding to the progeny of a fraction of 
the SGBb stellar population. 

Therefore, the IS of NGC$\,$1851 belongs to a second generation (SG) of stars.  
That a SG exists within the SGBb component is 
in agreement with recent spectroscopy of the SGB stars,
showing that each SGB hosts multiple subgenerations of 
stars \citep{lardo12}.
\citet{dantona08} have also postulated that the longer periods of the NGC$\,$1851
RR Lyrae may indicate that these variables 
may belong to the SG, and our synthetic HB strengthens this notion.  
Chemical anomalies in GCs suggest that Ôself-enrichmentÕ is a common 
feature among GCs.
The quasi-constancy of heavy metals in most GCs leads to the assumption that abundance 
variations are not or scarcely affected by SN ejecta, but involves formation of a second generation 
of stars from matter processed into the FG stars.  

The SG will most probably show a spread in He \citep{dantona08} 
because the self-enriched material may come from different progenitors that have different 
chemical peculiarities, or may be diluted in different fractions with matter from the first generation (FG).  
We remark that such a helium spread is an essential ingredient in order to reproduce the 
pulsation properties of the RR Lyrae population as a whole.
Simulations using constant He across the IS give synthetic period distributions
that do not match the observed one as well and result in the observed $\rm N_1/N_{tot}$
being lower than what is observed.  
Actually, one can note that variations of He in 19 GCs have 
also been deduced by \citet{bragaglia10} from 1400 RGB stars with abundance determinations.
As discussed in \citet{gratton10}, a star-to-star spread in the He abundance may explain many
aspects of the horizontal branches of GCs.  

It is worth pointing out that there have been suggestions 
of problems in the late stages of HB evolution in current HB tracks
\citep{catelan09b,valcarce08}.  Moreover, \citet{catelan09b} show 
that in the case of NGC$\,$1851, over-luminous stars on the blue HB could be interpreted 
by an underestimate of the luminosity evolution along the HB rather than in terms of a 
moderate level of helium enrichment.  Here we do not attempt to resolve this ambiguity 
for NGC$\,$1851; rather, we assume that the evolutionary tracks adopted  
represent the HB evolution accurately, and remind the reader that our comparisons are 
ultimately subject to both theoretical and observational uncertainties.

The pulsation periods and 
amplitudes from the RR Lyrae variables resulting from variations in He along the IS
have different characteristics.  The RR Lyrae variables with a ``normal" helium
have periods and amplitudes, as well as a $N_{\rm 1}/N_{\rm tot}$
ratio, that is inline with OoI-type GCs.  In contrast, the RR Lyrae variables with slightly enhanced
He (0.27 $<$ $Y$ $<$ 0.28) have longer periods and a higher ratio of $N_{\rm 1}/N_{\rm tot}$, 
indicative of RR Lyrae variables in OoII-type GCs.  In the absence of spectroscopy of the RR Lyrae variables
in NGC$\,$1851, the synthetic horizontal part of the HB and RR Lyrae instability strip 
presented here is the simplest one that reproduces the 
available observations with the smallest amount of free parameters.  New observations
of the RR Lyrae variables may require more complex modeling, however, and would be
particularly interesting.

Oosterhoff-I clusters tend to be more metal-rich and 
host fainter RR Lyrae variables than OoII clusters \citep{caputo00}.  
As the metallicity has an effect on the absolute magnitude of an RR Lyrae, 
it has been difficult to disentangle whether the metallicity difference alone is affecting 
the brightness differences, or whether there are differences in the intrinsic 
magnitudes of RR Lyrae variables in OoI and OoII
globular clusters caused by something other than just metallicity (like evolution
or helium).  In this cluster, where an internal spread in $\rm [Fe/H]$ is small at most, 
our results indicate that a difference in helium abundance in the RR Lyrae
variables is affecting where in the PA diagram the RR Lyrae star falls.

Our model consists of a BHB that is He-enriched ($Y \sim$0.28) yet older than the RHB ($Y$=0.248).
This can be explained if the cluster was formed by a process such 
as a merger with populations that differ in He and age.
Such a scenario has already been discussed by e.g. \citet{carretta11b} and \citet{bekki12}.  
Hence the BHB would not be a second generation (SG) of stars originating from
the same population as the RHB (the SGBb).

Since this paper was submitted, results from an intermediate resolution spectroscopic analysis 
of the two SGBs by \citet{gratton12new} indicate
that the $\rm [Fe/H]$ difference between the SGBb and SGBf is $\sim$0.07 dex, the SGBf being 
more metal rich.  The RR Lyrae stars in our scenario are the progeny of
SGBb; hence this metallicity difference does not affect our results.  We find that for BHB stars hotter 
than the IS a 0.1~dex increase in $\rm [Fe/H]$ at fixed $\rm (V-I)$ changes the HB masses 
(at fixed $Y$) by $\sim$0.01$M_{\sun}$, and $\rm M_V$ changes by $\sim$0.01 mag.
Therefore the effect of such a $\rm [Fe/H]$ difference between the two SGBs has a negligible
effect on our HB modeling.

We have shown that a spread in 
He reproduces the pulsational properties of the RR Lyrae sample as a whole, indicating the 
presence of a SG of stars in NGC$\,$1851.  Our analysis therefore demonstrates 
that RR Lyrae properties in a given GC can provide suitable constraints on the multiple 
population phenomenon in that GC.  It is worth carrying out more studies of this kind
to investigate further this connection with the occurrence of the multiple population phenomenon, 
especially in GCs with a sizable population of RR Lyrae stars and in which the stellar chemical 
patterns are well known.

\acknowledgments
The authors thank Aaron Dotter for helpful discussions.
S.C warmly thanks PRIN INAF 2009 ``Formation and early evolution 
of massive star clusters" (P.I.: R. Gratton) and PRIN INAF 2011 ``Multiple populations in 
Globular Clusters: their role in the Galaxy assembly" (P.I.: E. Carretta)
for financial support.  Support for M.C. and P.A. is provided by
the Chilean Ministry for the Economy, Development, and Tourism's Programa
Iniciativa Cient\'{i}fica Milenio through grant P07-021-F, awarded to The
Milky Way Millennium Nucleus; by Proyecto Fondecyt Regular \#1110326;
by the BASAL Center for Astrophysics and Associated Technologies
(PFB-06); and by Proyecto Anillo ACT-86.
We would like to thank the anonymous referee whose thorough report 
has led to substantial improvements to this paper.

\clearpage

\begin{table}
\begin{scriptsize}
\centering
\caption{MW GCs with a population of RR Lyrae variables}
\label{GClist}
\begin{tabular}{lllllll} \hline
Name1 & Name2 & $\rm [Fe/H]$ & HB Type & Oo-Type & $\rm N_{RR0}$ & Reference \\ \hline
NGC 362 & - & $-$1.30 & $-$0.87 & OoI & 21 & \citet{szekely07} \\
NGC 3201 & - & $-$1.51 & +0.08 & OoI & 58 & \citet{piersimoni02, layden03} \\
NGC 4590 & M68 & $-$2.27 & +0.17 & OoII & 12 & \citet{walker94} \\
IC 4499 & - & $-$1.62 & +0.11 & OoI & 62 & \citet{kunder11a} \\
Ruprecht 106 & - & $-$1.78 & $-$0.82 & OoI & 12 & \citet{kaluzny95} \\
NGC 5053 & - & $-$2.30 & +0.52 & OoII & 6 & \citet{nemec04} \\
NGC 6934 & - & $-$1.56 & +0.25 & OoI & 51 & \citet{kaluzny01} \\
NGC 6981 & M72 & $-$1.48 & +0.14 & OoI & 20 & \citet{dickens72} \\
NGC 7006 & - & $-$1.46 & $-$0.28 & OoI & 43 & \citet{wehlau99} \\
NGC 7078 & M15 & $-$2.33 & +0.67 & OoII & 50 & \citet{corwin08} \\
NGC 5139 & wCen & $-$1.64 & +0.89 & OoII & 84 & \citet{kaluzny04, benko06} \\
NGC 5272 & M3 & $-$1.50 & +0.18 & OoI & 152 & \citet{corwin01} \\
NGC 5466 & - & $-$1.70 & +0.58 & OoII & 12 & \citet{corwin99} \\
NGC 6229 & - & $-$1.43 & +0.24 & OoI & 29 & \citet{borissova01} \\
NGC 6426 & - & $-$2.15 & +0.58 & OoII & 8 & \citet{papadakis00} \\ 
NGC 6441 & - & $-$0.44 & $-$0.73 & OoIII & 25 & \citet{pritzl03} \\
NGC 6388 & - & $-$0.45 & $-$0.69 & OoIII & 4 & \citet{pritzl02} \\
NGC 6266 & M62 & $-$1.18 & +0.55 & OoI & 74 & \citet{contreras10} \\
NGC 6333 & M9 & $-$1.79 & +0.87 & OoII & 6 & \citet{clement99b} \\
NGC 6341 & M92 & $-$2.35 & +0.91 & OoII & 11& \citet{kopacki01} \\
NGC 6362 & - & $-$1.07 & $-$0.58 & OoI & 19 & \citet{olech01} \\
NGC 6626 & M28 & $-$1.46 & +0.90 & OoI & 8 & \citet{wehlau90} \\
NGC 5904 & M5 & $-$1.33 & +0.31 & OoI & 52 & \citet{szeidl11} \\
NGC 5986 & - & $-$1.63 & +0.97 & OoII & 6 & \citet{alves01} \\
NGC 6121 & M4 & $-$1.18 & $-$0.06 & OoI  &15 & \citet{cacciari79, sturch77} \\
NGC 6171 & M107 & $-$1.03 & $-$0.73 & OoI  & 15 & \citet{clement97, dickens70} \\ 
NGC 6712 & - & $-$1.02 & $-$0.62 & OoI & 7 & \citet{sandage66} \\
NGC 6723 & - & $-$1.10 & $-$0.08 & OoI & 15 & \citet{menzies74} \\
NGC 6864 & M75 & $-$1.29 & $-$0.07 & OoI & 8 & \citet{corwin03}  \\
NGC 5024 & M53 & $-$2.06 & +0.81 & OoII & 22 & \citet{arellano11} \\
NGC 7089 & M2 & $-$1.66 & +0.92 & OoII & 18 & \citet{lee99a} \\
NGC 2419 & - & $-$2.20 & +0.86 & OoII & 28 & \citet{dicriscienzo11} \\
NGC 1851 & - & $-$1.18 & $-$0.32 & OoI & 19 & \citet{walker98} \\
NGC 5286 & - & $-$1.70 & +0.80 & OoII & 27 & \citet{zorotovic10} \\
NGC 2808 & - & $-$1.18 & $-$0.49 &  OoI & 6 & \citet{kunder11b} \\
NGC 6101 & - & $-$1.98 & -- & OoII  & 3 & \citet{cohen11} \\
NGC 4147 & - & $-$1.78 & +0.55 & OoI & 5 & \citet{stetson05} \\
NGC 6809 & M55 & $-$1.93 & -- & OoI & 4  & \citet{olech99} \\
NGC 6715 & M54 & $-$1.44 & +0.54 & OoI & 52 & \citet{sollima10} \\
\hline
\end{tabular}
 \end{scriptsize}
 \end{table}

\clearpage

\end{document}